\documentclass[journal,onecolumn]{IEEEtran}
\usepackage{amssymb}
\usepackage{amsmath}
\usepackage{amsthm}
\usepackage{bm}

\ifCLASSINFOpdf
  
\else
  
\fi

\begin{document}

\title{Codes over Finite Ring $\mathbb{Z}_k$, MacWilliams Identity and Theta Function}

\author{Zhiyong Zheng$^{3,1,2}$, Fengxia Liu$^{*1,2,3}$ and Kun Tian$^{*3,1,2}$
        % <-this % stops a space
\thanks{$^{1}$Great Bay University, Dongguan, 523830, China.}% <-this % stops a space
\thanks{$^{2}$Henan Academy of Sciences, Zhengzhou, 450046, China.}
\thanks{$^{3}$Engineering Research Center of Ministry of Education for Financial Computing and Digital Engineering, Renmin University of China, Beijing, 100872, China.}
\thanks{$^{*}$Corresponding author email: shunliliu@gbu.edu.cn, tkun19891208@ruc. edu.cn}}

\maketitle

% As a general rule, do not put math, special symbols or citations
% in the abstract or keywords.
\begin{abstract}
In this paper, we study linear codes over $\mathbb{Z}_k$ based on lattices and theta functions. We obtain the complete weight enumerators MacWilliams identity and the symmetrized weight enumerators MacWilliams identity based on the theory of theta function. We extend the main work by Bannai, Dougherty, Harada and Oura to the finite ring $\mathbb{Z}_k$ for any positive integer $k$ and present the complete weight enumerators MacWilliams identity in genus $g$. When $k=p$ is a prime number, we establish the relationship between the theta function of associated lattices over a cyclotomic field and the complete weight enumerators with Hamming weight of codes, which is an analogy of the results by G. Van der Geer and F. Hirzebruch since they showed the identity with the Lee weight enumerators.
\end{abstract}

% Note that keywords are not normally used for peerreview papers.

% For peer review papers, you can put extra information on the cover
% page as needed:
% \ifCLASSOPTIONpeerreview
% \begin{center} \bfseries EDICS Category: 3-BBND \end{center}
% \fi
%
% For peerreview papers, this IEEEtran command inserts a page break and
% creates the second title. It will be ignored for other modes.
\IEEEpeerreviewmaketitle

\section{Introduction}

\IEEEPARstart{T}{he} interplay between coding theory and lattice theory has been a subject of profound significance in information transmission for several decades. Since the seminal work which established explicit connections between binary codes and lattices, this synergy has catalyzed advancements across both fields. Lattices derived from codes inherit algebraic structures that preserve optimal packing densities, while codes constructed via lattice projections benefit from geometric insights. Over decades, researchers have systematically extended this correspondence to broader algebraic frameworks. These extensions not only deepen theoretical understanding but also enhance practical applications in cryptography, quantization, and network coding.

The foundational work about the relationship between binary codes and lattices was presented by Conway et al. \cite{08} in 1988. Bonnecaze et al. \cite{05} marked a pivotal shift by constructing lattices from linear codes over $\mathbb{Z}_4$, revealing properties analogous to those of binary codes. Subsequent efforts generalized this framework to rings $\mathbb{Z}_{2^k}$ for arbitrary $k$ in \cite{10}, enabling systematic explorations of self-dual codes and their related lattices. A significant milestone emerged with Bannai et al. \cite{02}, who established correspondences between Type II codes over $\mathbb{Z}_{2k}$ and even unimodular lattices. A construction of even unimodular lattices is given using Type II codes in their work. There are also many works about the self-dual or type II codes over $\mathbb{Z}_k$ when $k$ is the power of $2$ or an even number (cf. \cite{01,04,06,09,24,35,37,46,50}). For example, Harada \cite{12,13,14} provided some ways for constructing self-dual or Type II codes in high dimensions or proved the existence of extremal Type II codes in some special cases.

The MacWilliams theorem for linear codes over a finite field $F_q$ establishes an identity that relates the weight enumerators of a code to the weight enumerators of its dual code (cf. \cite{31}). It was also demonstrated that the MacWilliams identity admits numerous generalizations. The first direction is focusing on some special codes over a finite field, such as the MacWilliams identities for binary codes, convolutional codes, cyclic codes and so on (cf. \cite{03,28,30,33,38,43}). The second direction involves generalizing the finite field $F_q$ to some finite rings, such as $\mathbb{Z}_k$, the Galois rings and the Frobenius rings (cf. \cite{27,47,48}). For instance, in \cite{26}, Klemm generalized the MacWilliams identity for codes defined over the finite rings $\mathbb{Z}_k$ in 1987. Some other works could be found in \cite{36,41}. The third direction involves generalizing the weight enumerators to include more than two variables, such as the Lee and complete weight enumerators for codes over a finite field or a finite ring (cf. \cite{07,23,29,45,49}). The first complete weight enumerators identity for codes over a finite field was provided by MacWilliams  \cite{32} in 1972. Wan \cite{44} provided the complete weight enumerators MacWilliams identity for codes over the Galois rings and Siap \cite{40} proposed that for codes over the matrix rings. Zheng et al. \cite{52} gave the complete weight enumerators MacWilliams identity for codes over $\mathbb{Z}_k^n[\xi]$ with a positive integer $k$ and a root $\xi$ of an irreducible polynomial.

Despite these achievements, critical limitations persisted. Previous results on the complete MacWilliams identities for codes over $\mathbb{Z}_k$ are mostly focused on special cases. For example, Bannai et al. \cite{02} only gave the complete weight enumerators MacWilliams identity in genus $g$ for codes over $\mathbb{Z}_k$ when $k$ is an even number. Hirzebruch only showed that the MacWilliams identity with the Lee weight enumerator when $k=p$ is a prime number based on the ring of algebraic integers over a cyclotomic field (cf. \cite{15,19,20}). The case of an arbitrary positive integer $k$ remained unresolved. These lead us to strong limitations in previous approaches that relied on parity constraints.

In this paper, we fill these gaps through a unified framework for linear codes over $\mathbb{Z}_k$ with an arbitrary positive integer $k$. Our contributions are multifaceted. By defining $k$ auxiliary theta functions, we establish an explicit relation between the theta function of the lattice associated with code $C\subset \mathbb{Z}_k^n$ and the complete weight enumerators of $C$ (Theorem 1), from which the complete MacWilliams identity arises naturally. This is a modification of the result by F. Hirzebruch since he constructed the identity with the Lee weight enumerators (cf. \cite{15,19,20}). We derive the complete weight enumerators and the symmetrized weight enumerators MacWilliams identity for codes over $\mathbb{Z}_k$ using the theory of theta functions (Theorem 2), thereby providing a novel interpretation in comparison with traditional proofs. We establish the complete weight enumerator MacWilliams identity in genus $g$ (Theorem 3) which is valid for all positive integers $k$. This significantly extends Theorem 5.1 of Bannai et al. in \cite{02}, which required $k$ to be even. This result not only unifies fragmented results of previous works but also provides more deep relations between codes, lattices, and theta functions. We give the relationship between the complete weight enumerators with Hamming weight of codes in $F_p^n$ and the theta function of associated lattices over a cyclotomic field (Theorem 4), which is a generalization of the results by G. Van der Geer and F. Hirzebruch since they showed the identity with the Lee weight enumerator.

%At the core of our methodology lies the interplay between complete weight enumerators and theta functions. By reformulating theta functions as generating series for code weights, we exploit the complete and  symmetrized weight enumerators MacWilliams identities.

\subsection{The Complete Weight Enumerators MacWilliams Identity}

The first complete weight enumerators identity was given by MacWilliams in \cite{32}. Let $F_q$ be a finite field of $q$ elements $a_0,a_1,\cdots,a_{q-1}$, $n$ be a positive integer, $C\subset F_q^n$ be a linear code, and $|C|$ be the number of codewords in $C$. We denote the complete weight enumerators of $C$ is
\begin{equation*}
W_C(X_0,X_1,\cdots,X_{q-1})=\sum\limits_{c\in C} X_0^{w_0(c)} X_1^{w_1(c)}\cdots X_{q-1}^{w_{q-1}(c)},
\end{equation*}
where $w_j(c)$ is the number of elements in the codeword $c$ which are equal to $a_j$, $0\leqslant j\leqslant q-1$. Here we focus on the case of the finite ring $\mathbb{Z}_k$ instead of the finite field $F_q$. In our previous work \cite{52}, we prove the complete weight enumerators MacWilliams identity for codes over $\mathbb{Z}_k$. If $C\subset \mathbb{Z}_k^n$ is a $k$-ary linear code, then we have
\begin{equation*}
W_{C^{\bot}}(X_0,X_1,\cdots,X_{k-1})=\frac{1}{|C|} W_C(\sum\limits_{j=0}^{k-1} X_j, \sum\limits_{j=0}^{k-1} e^{\frac{2\pi j}{k}i}X_j, \cdots, \sum\limits_{j=0}^{k-1} e^{\frac{2\pi(k-1) j}{k}i}X_j).
\end{equation*}
The general form of the above complete weight enumerators MacWilliams identity for codes over $R=\mathbb{Z}_k[\xi]$ also holds (cf. \cite{52}), where $\xi$ is a root of an irreducible polynomial.

From construction A, one can establish the correspondence between codes and lattices, which reveals the relationship between theta function, modular form and MacWilliams identity, such as Proposition 2.11 of \cite{11} and Gleason Theorem. \cite{02} generalized construction A for an even number $k$ and presented the correspondence between type II codes over $\mathbb{Z}_{2k}$ and even unimodular lattices, as well as the complete weight enumerators and symmetrized weight enumerators MacWilliams identities in genus $g$. In this paper, we extend construction A for any positive integer $k$ and provide the connections between a $k$-ary code $C$ and the associated lattice $\Gamma_C$. In particular, we prove that the theta function of $\Gamma_C$ could be expressed by the complete weight enumerators of $k$ theta functions $A_0(z),A_1(z),\cdots,A_{k-1}(z)$ as the following Theorem 1, which is a natural generalization of Proposition 2.11 in \cite{11}.

\textbf{Theorem 1} Let $C\subset \mathbb{Z}_k^n$ be a $k$-ary code, $\Gamma_C=\frac{1}{\sqrt{k}}\rho^{-1}(C)$ be the associated lattice of $C$, then
\begin{equation*}
\vartheta_{\Gamma_C}(z)=W_C(A_0(z),A_1(z),\cdots,A_{k-1}(z)).
\end{equation*}
Based on the result of Theorem 1, we could obtain the complete weight enumerators MacWilliams identity expressed by theta functions for any general modulus $k$.

\textbf{Theorem 2} Let $C\subset \mathbb{Z}_k^n$ be a $k$-ary code. Then we have
\begin{equation*}
W_{C^{\bot}}(A_0(z),A_1(z),\cdots,A_{k-1}(z))=\frac{1}{|C|} W_C(\sum\limits_{j=0}^{k-1} A_j(z), \sum\limits_{j=0}^{k-1} e^{\frac{2\pi j}{k}i}A_j(z), \cdots, \sum\limits_{j=0}^{k-1} e^{\frac{2\pi(k-1) j}{k}i}A_j(z)).
\end{equation*}

From the result of Theorem 2, one could get the symmetrized weight enumerators MacWilliams identity directly.

\subsection{The Complete Weight Enumerators MacWilliams Identity in Genus $g$}

Many researchers concentrated on high dimensional MacWilliams identities. For example, Kaplan introduced the $m$-tuple weight enumerators and proved the corresponding MacWilliams identity in \cite{25}. However, these results hold by codes defined over a finite field instead of a finite ring. Bannai et al. \cite{02} gave the complete weight enumerators MacWilliams identity in Genus $g$ for codes over $\mathbb{Z}_k$ and $k$ is an even number. For a $k$-ary code $C\subset \mathbb{Z}_k^n$, an even number $k$ and a positive integer $g$, Bannai et al. described the complete weight enumerators MacWilliams identity in genus $g$ is
\begin{equation*}
\mathfrak{C}_{C,g}(z_a\ \text{with}\ a\in \mathbb{Z}_k^g)=\sum\limits_{c_1,\cdots,c_g\in C} \mathop{\prod}\limits_{a\in \mathbb{Z}_k^g} z_a^{w_a(c_1,\cdots,c_g)},
\end{equation*}
by writing $c_1=(c_{11},c_{12},\cdots,c_{1n}),\cdots,c_g=(c_{g1},c_{g2},\cdots,c_{gn})$, here $w_a(c_1,\cdots,c_g)$ denotes
\begin{equation*}
w_a(c_1,\cdots,c_g)=\#\{i\ |\ (c_{1i},c_{2i},\cdots,c_{gi})=a,\ 1\leqslant i\leqslant n\}.
\end{equation*}
Suppose $f\in \mathbb{C}(x_1,x_2,\cdots,x_n)$ is a complex polynomial of $x_1,x_2,\cdots,x_n$, and $M$ is a matrix $(a_{ij})_{n\times n}$ of order $n$. We define $Mf(x_1,x_2,\cdots,x_n)$ by
\begin{equation*}
Mf(x_1,x_2,\cdots,x_n)=f\left(\sum\limits_{j=1}^n a_{1j}x_j,\sum\limits_{j=1}^n a_{2j}x_j,\cdots,\sum\limits_{j=1}^n a_{nj}x_j\right).
\end{equation*}
Bannai et al. established the following MacWilliams identity as Theorem 5.1 in \cite{02},
\begin{equation*}
\mathfrak{C}_{C^{\bot},g}(z_a)=\frac{1}{|C|^g} T\mathfrak{C}_{C,g}(z_a).
\end{equation*}
here $T=(\eta^{a\cdot b})_{a,b\in \mathbb{Z}_k^g}$ is a matrix of order $k^g$, and $\eta=e^{\frac{2\pi i}{k}}$ is the primitive root of unit. However, they only prove that it is right when $k$ is an even number. In this paper, we present the complete weight enumerators MacWilliams identity in genus $g$ for any positive integer $k$ as the following Theorem 3.

\textbf{Theorem 3} Let $C\subset \mathbb{Z}_k^n$ be a $k$-ary code. $\mathfrak{C}_{C,g}(z_a)$ is the complete weight enumerators in genus $g$. Then we have
\begin{equation*}
\mathfrak{C}_{C^{\bot},g}(z_a)=\frac{1}{|C|^g} T\mathfrak{C}_{C,g}(z_a).
\end{equation*}
Theorem 3 has the same form as the result of Bannai et al., while it holds for codes in $\mathbb{Z}_k^n$ with any positive integer $k$. We will prove it based on the Fourier transform and Poisson summation formula in Section IV. If $g=1$, the above result becomes the ordinary complete weight enumerators MacWilliams identity. We can also obtain the symmetrized weight enumerators MacWilliams identity in genus $g$ from Theorem 3 directly by treating $z_a$ and $-z_{a}$ as the equivalent elements in the complete weight enumerators.

\subsection{The Complete Weight Enumerators MacWilliams Identity in Cyclotomic Fields}

When $k=p$ is an odd prime number, associating a code over $F_p$ with lattice over cyclotomic field is due to G. van der Geer and F. Hirzebruch \cite{20} (also see \cite{11}, Chapter 5). They considered the Lee weight enumerators $S_C(X_0,X_1,\cdots,X_{\frac{p-1}{2}})$ for codes $C\subset F_p^n$ defined as the following
\begin{equation*}
S_C(X_0,X_1,\cdots,X_{\frac{p-1}{2}})=\sum\limits_{c\in C} X_0^{w_0(c)} X_1^{w_1(c)}\cdots X_{\frac{p-1}{2}}^{w_{\frac{p-1}{2}}(c)},
\end{equation*}
where $w_j(c)$ is the number of elements in the codeword $c$ which are equal to $a_j$ or $p-a_j$, $0\leqslant j\leqslant \frac{p-1}{2}$. Van der Geer and Hirzebruch provided the Alpbach Theorem which established the MacWilliams identity between the Lee weight enumerators and theta function (see \cite{11}, Theorem 5.3),
\begin{equation*}
\theta_{\Gamma_C}(z)=S_C(\theta_0(z),\theta_1(z),\cdots,\theta_{\frac{p-1}{2}}(z)),
\end{equation*}
here $\theta_0(z),\theta_1(z),\cdots,\theta_{\frac{p-1}{2}}(z)$ are $\frac{p+1}{2}$ theta functions, which is about half of the number of elements in $F_p$. In this paper, we generalize their results to complete weight enumerators by defining $p$ theta functions $\vartheta_0(z),\vartheta_1(z),\cdots,\vartheta_{p-1}(z)$, and present the identity to show the relationship between the complete weight enumerators with Hamming weight of codes in $F_p^n$ and the theta function of associated lattices over a cyclotomic field in Theorem 4.

To state our results, we define the following notations. Let $\xi=e^{\frac{2\pi i}{p}}$, $K=\mathbb{Q}(\xi)$ be the cyclotomic field, $K^+=\mathbb{Q}(\xi+\xi^{-1})$ be the maximal real subfield of $K$, $\text{Tr}_{K/\mathbb{Q}}$ and $\text{Tr}_{K^+/\mathbb{Q}}$ be the trace respectively, $\mathfrak{D}$ be the integers ring of $K$, $\mathfrak{B}=\left<1-\xi\right>$ be the principal ideal of $\mathfrak{D}$ generated by the element $1-\xi\in \mathfrak{D}$.

To associate a linear code $C\subset F_p^n$ and a lattice $\Gamma_C\subset \mathfrak{D}^n$, suppose $C\subset C^{\bot}$, and $\rho:\mathfrak{D}^n\longrightarrow (\mathfrak{D}/\mathfrak{B})^n$ is the mapping defined by the reduction modulo the principal ideal $\mathcal{B}$ in each coordinate. We define the lattice and theta function by
\begin{equation*}
\Gamma_C=\rho^{-1}(C)\subset \mathfrak{D}^n,
\end{equation*}
\begin{equation*}
\vartheta_{\Gamma_C}(z)=\sum\limits_{x\in \Gamma_C}e^{2\pi iz \text{Tr}_{K^+/\mathbb{Q}}(\frac{x \overline{x}}{p})},
\end{equation*}
where $z\in \mathbb{H}$ is the upper half plane of complex number.

For any $j=0,1,2,\cdots,p-1$, let
\begin{equation*}
\vartheta_j(z)=\sum\limits_{x\in \mathfrak{B}+j} e^{2\pi iz \text{Tr}_{K^+/\mathbb{Q}}(\frac{x \overline{x}}{p})}.
\end{equation*}
Our main result is the following Theorem 4.

\textbf{Theorem 4} Let $C\subset F_p^n$ be a linear code such that $C\subset C^{\bot}$. $W_C(X_0,X_1,\cdots,X_{p-1})$ is the complete weight enumerator of $C$ with Hamming weight, then we have
\begin{equation*}
\vartheta_{\Gamma_C}(z)=W_C(\vartheta_0(z),\vartheta_1(z),\cdots,\vartheta_{p-1}(z)).
\end{equation*}

Van der Geer and Hirzebruch gave their Alpbach Theorem (Theorem 5.3 of \cite{11}) with Lee weight enumerator of $C$ in a higher dimensional form. There is a similar relation between the complete weight enumerator of a linear code $C\subset F_p^n$ with $C\subset C^{\bot}$ and certain Jacobi forms over the field $\mathbb{Q}(\xi+\xi^{-1})$ in \cite{a}.

The remainder of this paper is structured as follows. Section II reviews preliminaries: the correspondences between codes and lattices, construction of type II codes, and the properties of theta functions. Section III presents the complete weight enumerators MacWilliams identity for codes over the finite ring $\mathbb{Z}_k$ for any positive integer $k$ (Theorem 2) based on theta functions, with the symmetrized weight enumerators MacWilliams identity as a corollary. Section IV generalizes the complete weight enumerators MacWilliams identity to genus $g$ (Theorem 3), accompanied the proof by Fourier transform and Poisson summation formula. Section V gives the complete weight enumerators MacWilliams identity in cyclotomic fields (Theorem 4). Section VI concludes our results with open questions.

\section{Preliminaries}

The relationship between codes and lattices have been studied for a few decades. In \cite{02}, Bannai et al. introduce Type II codes $C$ over $\mathbb{Z}_{2k}$ which are closely related to even unimodular lattices. They use the Euclidean weight $wt_E(c)=\sum\limits_{i=1}^n \min\{c_i^2,(2k-c_i)^2\}$ as a norm for a codeword $c=(c_1,c_2,\cdots,c_n)\in C\subset \mathbb{Z}_{2k}^n$, and define a Type II code over $\mathbb{Z}_{2k}$ as a self-dual code with Euclidean weights divisible by $4k$. They prove that $C$ is type II if and only if the lattice associated by $C$ is an even unimodular lattice. In this section, we modify the Euclidean weight of a code word $c=(c_1,c_2,\cdots,c_n)$ by $wt_E(c)=\sum\limits_{i=1}^n c_i^2$ for convenience and get the type II codes more directly. First let us introduce some definitions for codes and lattices.

Let $\mathbb{Z}_k$ be the ring of integers modulo $k$ with positive integer $k>1$. Consider the following reduction mod $k$
\begin{equation*}
\rho: \mathbb{Z}^n\longrightarrow (\mathbb{Z}/k\mathbb{Z})^n=\mathbb{Z}_k^n,
\end{equation*}
it is easy to see that this is a homomorphism. A code $C$ of length $n$ over the ring $\mathbb{Z}_k$ is a subset of $\mathbb{Z}_k^n$, and if the code is an additive subgroup of $\mathbb{Z}_k^n$ then it is a linear code. Unless otherwise stated all codes will be linear. For any $c=(c_1,c_2,\cdots,c_n)\in C$, we call it a $k$-ary codeword with length $n$.
The preimage of $C$ in $\mathbb{Z}^n$ is denoted by $\rho^{-1}(C)$, and it is a subgroup of $\mathbb{Z}^n$. Therefore, $\rho^{-1}(C)$ is a lattice in $\mathbb{R}^n$.

\textbf{Definition 1} Let $C\subset \mathbb{Z}_k^n$ be a $k$-ary code. The lattice associated with the code $C$ is defined as
\begin{equation*}
\Gamma_C=\frac{1}{\sqrt{k}}\rho^{-1}(C).
\end{equation*}
Equivalently, we can write $\Gamma_C$ as
\begin{equation*}
\Gamma_C=\frac{1}{\sqrt{k}}\{c+kz\ |\ c\in C, \text{and}\ z\in \mathbb{Z}^n\}.
\end{equation*}

\textbf{Definition 2} Let $C\subset \mathbb{Z}_k^n$ be a $k$-ary code. For any $c_1,c_2\in C$, let $c_1=(c_{11},c_{12},\cdots,c_{1n})$, $c_2=(c_{21},c_{22},\cdots,c_{2n})$. We define the inner product of $c_1$ and $c_2$ in $\mathbb{Z}_k$ as
\begin{equation*}
c_1\cdot c_2=\sum\limits_{i=1}^n c_{1i}c_{2i}\ \text{mod}\ k.
\end{equation*}

\textbf{Definition 3} (1) Let $C\subset \mathbb{Z}_k^n$ be a $k$-ary code, we define the dual code of $C$ as $C^{\bot}=\{c'\in \mathbb{Z}_k^n\ |\ c\cdot c'=0\ \text{for all}\ c\in C\}$.

(2) A linear code $C\subset \mathbb{Z}_k^n$ is called self-orthogonal if $C\subset C^{\bot}$, that is $c_1\cdot c_2=0$ for all $c_1,c_2\in C$.

(3) A linear code $C\subset \mathbb{Z}_k^n$ is called self-dual if $C=C^{\bot}$.

\textbf{Definition 4} If $k$ is an even number, a linear code $C\subset \mathbb{Z}_k^n$ is called doubly even if the Euclidean weight of any codeword $c=(c_1,c_2,\cdots,c_n)\in C$ satisfies that $c_1^2+c_2^2+\cdots+c_n^2\equiv 0\ (\text{mod}\ 2k)$. If a code is self-dual and doubly even, then it is called a type II code.

It's easy to see that if $c\equiv c'\ (\text{mod}\ k)$, $c=(c_1,c_2,\cdots,c_n)$, $c'=(c_1',c_2',\cdots,c_n')$, then
\begin{equation*}
c_1^2+c_2^2+\cdots+c_n^2\equiv c_1'^2+c_2'^2+\cdots+c_n'^2\ (\text{mod}\ 2k),
\end{equation*}
so the definition of doubly even code does not depend on the selection of the representative element in $\mathbb{Z}_k$ and it is well defined.

\textbf{Remark 1} In \cite{02}, Bannai et al. use the Euclidean weight $\sum\limits_{i=1}^n \min\{c_i^2,(k-c_i)^2\}$ for a codeword $c=(c_1,c_2,\cdots,c_n)\in C\subset \mathbb{Z}_{k}^n$ in the above definition when $k$ is even. Definition 4 of type II code is equivalent to that in \cite{02}. We will show that it is more convenient to construct a type II code by our definition in this section.

We refer to \cite{09,10,46} for any elementary facts about codes over finite rings. For example, \cite{09,10} shows that the numbers of codewords in $C$ and $C^{\bot}$ satisfy $|C|\cdot |C^{\bot}|=k^n$. \cite{10} proves that if $k$ is a square then there exist self-dual codes over $\mathbb{Z}_k$ for all lengths, as well as if $C$ is a self-dual code of odd length over $\mathbb{Z}_k$ then $k$ is a square.

\textbf{Definition 5} (1) Let $\Gamma\subset \mathbb{R}^n$ be a lattice, we define the dual lattice of $\Gamma$ as $\Gamma^*=\{y\in \mathbb{R}^n\ |\ x\cdot y\in \mathbb{Z}\ \text{for all}\ x\in \Gamma\}$.

(2) A lattice $\Gamma\subset \mathbb{R}^n$ is called integral if $\Gamma\subset \Gamma^*$, that is $x\cdot y\in \mathbb{Z}$ for all $x,y\in \Gamma$.

(3) A lattice $\Gamma\subset \mathbb{R}^n$ is called self-dual or unimodular if $\Gamma=\Gamma^*$.

(4) A lattice $\Gamma\subset \mathbb{R}^n$ is called even if $x^2$ is an even number for all $x\in \Gamma$.

The following Proposition 1 gives the correspondence between a code $C$ and the associated lattice $\Gamma_C$.

\textbf{Proposition 1} Let $C\subset \mathbb{Z}_k^n$ be a $k$-ary code and $\Gamma_C$ be the associated lattice of $C$. We have the following results:

(1) $C\subset C^{\bot}$ if and only if $\Gamma_C\subset \Gamma_C^*$.

(2) If $k$ is an even number, then $C$ is doubly even if and only if $\Gamma_C$ is an even lattice.

(3) $C$ is self-dual if and only if $\Gamma_C$ is unimodular.

\textbf{Proof:} (1) For any $x,y\in \Gamma_C$, we have $x=\frac{1}{\sqrt{k}}(c_1+kz_1)$, $y=\frac{1}{\sqrt{k}}(c_2+kz_2)$, here $c_1,c_2\in C$ and $z_1,z_2\in \mathbb{Z}^n$. Then we have
\begin{equation*}
x\cdot y=\frac{1}{k}(c_1 c_2+kc_1 z_2+kc_2 z_1+k^2 z_1 z_2)\equiv \frac{1}{k}c_1 c_2\ (\text{mod}\ \mathbb{Z}),
\end{equation*}
it follows that $x\cdot y\in \mathbb{Z}$ for all $x,y\in \Gamma_C$ if and only if $c_1 c_2=0$ in $\mathbb{Z}_k$ for all $c_1,c_2\in C$. Therefore, $\Gamma_C\subset \Gamma_C^*$ if and only if $C\subset C^{\bot}$.

(2) For any $x\in \Gamma_C$, we have $x=\frac{1}{\sqrt{k}}(c+kz)$, here $c\in C$ and $z\in \mathbb{Z}^n$. Since $k$ is an even number, then
\begin{equation*}
x^2=\frac{1}{k}(c^2+2kcz+k^2 z^2)=\frac{1}{k}c^2+2cz+k z^2\equiv \frac{1}{k}c^2\ (\text{mod}\ 2\mathbb{Z}),
\end{equation*}
it follows that $x^2\in 2\mathbb{Z}$ for all $x\in \Gamma_C$ if and only if $c^2\in 2k\mathbb{Z}$ for all $c\in C$. This means that $\Gamma_C$ is even if and only if $C$ is doubly even.

(3) If $C=C^{\bot}$, we know that $|C|=k^{\frac{n}{2}}$. Since $\mathbb{Z}^n/\rho^{-1}(C)\cong \mathbb{Z}_k^n/C$, we have
\begin{equation*}
|\mathbb{Z}^n/\rho^{-1}(C)|=|\mathbb{Z}_k^n/C|=k^{n-\frac{n}{2}}=k^{\frac{n}{2}},  \tag{2.1}
\end{equation*}
one can get
\begin{equation*}
\text{det}(\rho^{-1}(C))=\text{vol}(\mathbb{R}^n/\rho^{-1}(C))=|\mathbb{Z}^n/\rho^{-1}(C)|\text{vol}(\mathbb{R}^n/\mathbb{Z}^n)=k^{\frac{n}{2}},  \tag{2.2}
\end{equation*}
then
\begin{equation*}
\text{det}(\Gamma_C)=\frac{\text{det}(\rho^{-1}(C))}{k^{\frac{n}{2}}}=1\Rightarrow \text{det}(\Gamma_C)=\text{det}(\Gamma_C^*)=1.  \tag{2.3}
\end{equation*}
Based on the result of (1), we can get $\Gamma_C\subset \Gamma_C^*$ due to $C\subset C^{\bot}$, combine with (2.3) it follows that $\Gamma_C=\Gamma_C^*$.

On the other hand, if $\Gamma_C=\Gamma_C^*$, then $\text{det}(\Gamma_C)=1$. From (2.1), (2.2) and (2.3) we know $\text{det}(\rho^{-1}(C))=k^{\frac{n}{2}}$ and $|C|=|C^{\bot}|=k^{\frac{n}{2}}$. Based on (1), we have $C\subset C^{\bot}$ according to $\Gamma_C\subset \Gamma_C^*$, hence $C=C^{\bot}$ since the numbers of codewords of $C$ and $C^{\bot}$ are the same. Therefore, $C$ is self-dual if and only if $\Gamma_C$ is unimodular.\\
\hspace*{17.7cm} $\Box$

We know that there exists a Type II code of length $n$ over $\mathbb{Z}_k$ when $k$ is even if and only if $n$ is a multiple of eight \cite{02}, and Bannai et al. give the example of type II code in $Z_k^8$ for any even number $k$. Here we construct an example more conveniently with our modified Euclidean weight to show the existence of type II code on $\mathbb{Z}_k^{8n}$ with the length of any multiple of $8$ when $k$ is even.

\textbf{Example 1} For any even number $k$, from Lagrange's theorem on sums of squares, there are elements $a,b,c,d$ in $\mathbb{Z}_k$ such that
\begin{equation*}
1+a^2+b^2+c^2+d^2=2k.
\end{equation*}
Let $M$ be the following matrix
\begin{equation*}
M=\left(\begin{array}{cccc} a & b & c & d\\ b & -a & -d & c\\c & d & -a & -b\\d & -c & b & -a \end{array}\right).
\end{equation*}
We denote $I_{4n}$ by the identity matrix of order $4n$ and $M_{4n}$ by the block matrix of order $4n$ composed of $M$ in the diagonal
\begin{equation*}
M_{4n}=\left(\begin{array}{cccc} M & & & \\ & M & & \\ & & \ddots & \\ & & & M \end{array}\right).
\end{equation*}
Then the matrix $G=(I_{4n},M_{4n})$ generates a Type II code $C$ of length $8n$ over $\mathbb{Z}_k$.

\textbf{Proof:} First we show that $C$ is doubly even. Let the rows of matrix $G$ be $e_1,e_2,\cdots,e_{4n}$. For any $c\in C$, it can be written as $c=k_1 e_1+k_2 e_2+\cdots+k_{4n} e_{4n}$, $k_1,k_2,\cdots,k_{4n}\in \mathbb{Z}_k$. Note that $e_i^2$ is divisible by $2k$ and $e_i\cdot e_j$ is divisible by $k$ for $i\neq j$ in $\mathbb{Z}$, therefore,
\begin{equation*}
c^2\equiv(k_1 e_1+k_2 e_2+\cdots+k_{4n} e_{4n})^2=\sum\limits_{i=1}^{4n} k_i^2 e_i^2+\sum\limits_{1\leqslant j<l\leqslant 4n} 2k_j k_l e_j e_l\equiv 0\ (\text{mod}\ 2k),
\end{equation*}
It follows that $C$ is doubly even. Next we prove $C$ is self-dual. For any $c_1,c_2\in C$, assume $c_1=a_1 e_1+a_2 e_2+\cdots+a_{4n} e_{4n}$ and $c_2=b_1 e_1+b_2 e_2+\cdots+b_{4n} e_{4n}$, here $a_i,b_i\in \mathbb{Z}_k$ for $1\leqslant i\leqslant 4n$. It is easy to compute that
\begin{equation*}
c_1\cdot c_2\equiv(a_1 e_1+a_2 e_2+\cdots+a_{4n} e_{4n})(b_1 e_1+b_2 e_2+\cdots+b_{4n} e_{4n})\equiv 0\ (\text{mod}\ k).
\end{equation*}
Hence, we have $c_1\cdot c_2=0$ in $\mathbb{Z}_k$ for any $c_1,c_2\in C$, this means that $C\subset C^{\bot}$. Since $C$ is generated by the matrix $G$ and the $4n$ rows of $G$ are linearly independent in $\mathbb{Z}_k$, it follows that the number of codewords in $C$ is $|C|=k^{4n}$. Based on $|C|\cdot |C^{\bot}|=k^{8n}$ we have $|C^{\bot}|=k^{4n}$, i.e. $|C|=|C^{\bot}|=k^{4n}$. Combine with $C\subset C^{\bot}$, one can get $C=C^{\bot}$.

So we have proved that $C$ is a doubly even and self-dual code.\\
\hspace*{17.7cm} $\Box$

As another part of the preliminaries, let's introduce the theta function and modular form. The detailed contents could be found in \cite{b} by Shi, Choie, Sharma and Sol\'{e}. Suppose $k$ is an even number and $C\subset \mathbb{Z}_k^n$ is a $k$-ary doubly even and self-dual code. In order to show the property of modular form for theta function of $\Gamma_C$, let's begin with some definitions. We denote the group
\begin{equation*}
SL_2(\mathbb{Z})=\left\{g=\begin{pmatrix} a\ \ b\\ c\ \ d \end{pmatrix}\ \Bigg|\ a,b,c,d\in \mathbb{Z},\ ad-bc=1\right\}.
\end{equation*}
For any $g=\begin{pmatrix} a\ \ b\\ c\ \ d \end{pmatrix}\in SL_2(\mathbb{Z})$, we define $g(z)=\frac{az+b}{cz+d}$ as a function of $z$. Let $S$ and $T$ be the elements of $G=SL_2(\mathbb{Z})/\{\pm 1\}$ as
\begin{equation*}
S=\begin{pmatrix} 0\ -1\\ 1\ \ \ 0 \end{pmatrix}\ \ \text{and}\ \ T=\begin{pmatrix} 1\ \ 1\\ 0\ \ 1 \end{pmatrix},
\end{equation*}
therefore, $S(z)=-\frac{1}{z}$ and $T(z)=z+1$. The group $G$ is generated by $S$ and $T$ and the detailed proof could be found in \cite{34}.

Suppose $\mathbb{H}$ is the upper half plane
\begin{equation*}
\mathbb{H}=\{z\in \mathbb{C}\ |\ \text{Im}z>0\}\subset \mathbb{C}.
\end{equation*}
For $z\in \mathbb{H}$, let $t=e^{k\pi iz}$, $q=e^{2\pi iz}$ as usual. The classical theta function is written as $\vartheta_{\Gamma}(z)=\sum\limits_{x\in \Gamma} q^{\frac{1}{2}x\cdot x}$, while for the convenience of the results in the next section, we write the theta function $\vartheta_{\Gamma}(z)$ in a new form as $\vartheta_{\Gamma}(z)=\sum\limits_{x\in \Gamma} t^{\frac{1}{k}x\cdot x}$ since
\begin{equation*}
\vartheta_{\Gamma}(z)=\sum\limits_{x\in \Gamma} t^{\frac{1}{k}x\cdot x}=\sum\limits_{x\in \Gamma}e^{\pi i z x^2}=\sum\limits_{x\in \Gamma}q^{\frac{1}{2}x\cdot x}.  \tag{2.4}
\end{equation*}

\textbf{Definition 6} Let $m$ be an even number. A holomorphic function $f: \mathbb{H}\longrightarrow \mathbb{C}$ is called a modular form of weight $m$, if the following two conditions are satisfied:

(1) For any $\begin{pmatrix} a\ \ b\\ c\ \ d \end{pmatrix}\in SL_2(\mathbb{Z})$,
\begin{equation*}
f\left(\frac{az+b}{cz+d}\right)=(cz+d)^m f(z).
\end{equation*}

(2) $f$ is holomorphic at $z=i\infty$.

The following Proposition 2 shows the property of modular form for lattices associated with doubly even and self-dual codes on $\mathbb{Z}_k^n$ when $k$ is even.

\textbf{Proposition 2} If $k$ is even, and $C$ is a doubly even and self-dual code in $\mathbb{Z}_k^n$. Then we have $n\equiv 0\ (\text{mod}\ 8)$ and $\vartheta_{\Gamma_C}$ is a modular form of weight $\frac{n}{2}$.

Proposition 2 is a natural generalization of the result of modular form for binary doubly even and self-dual codes in \cite{11}. We claim that it also holds for $k$-ary doubly even and self-dual codes when $k$ is even. In order to prove Proposition 2, we give the following proposition first, which constructs the identity between the theta functions of $\Gamma_C$ and $\Gamma_C^*$.

\textbf{Proposition 3} Let $C\subset \mathbb{Z}_k^n$ be a $k$-ary code, we have
\begin{equation*}
\vartheta_{\Gamma_C}\left(-\frac{1}{z} \right)=\frac{1}{\text{det}(\Gamma_C)} \left(\frac{z}{i}\right)^{\frac{n}{2}} \vartheta_{\Gamma_C^*}(z).
\end{equation*}

\textbf{Proof:} According to the definition of theta function,
\begin{equation*}
\vartheta_{\Gamma_C}\left(-\frac{1}{z} \right)=\sum\limits_{x\in \Gamma_C} e^{-\frac{\pi i x^2}{z}}.
\end{equation*}
After some calculation we obtain the Fourier transform of $e^{-\frac{\pi i x^2}{z}}$ is $\left(\frac{z}{i}\right)^{\frac{n}{2}} e^{\pi i z x^2}$. Based on the Poisson summation formula,
\begin{equation*}
\vartheta_{\Gamma_C}\left(-\frac{1}{z} \right)=\sum\limits_{x\in \Gamma_C} e^{-\frac{\pi i x^2}{z}}=\frac{1}{\text{det}(\Gamma_C)} \sum\limits_{x\in \Gamma_C^*} \left(\frac{z}{i}\right)^{\frac{n}{2}} e^{\pi i z x^2}=\frac{1}{\text{det}(\Gamma_C)} \left(\frac{z}{i}\right)^{\frac{n}{2}} \vartheta_{\Gamma_C^*}(z).
\end{equation*}
\hspace*{17.7cm} $\Box$

\textbf{Proof of Proposition 2:} From Proposition 1 we know $\Gamma_C$ is an even and unimodular lattice. First we show that $n\equiv 0\ (\text{mod}\ 8)$. Suppose that $n$ is not divisible by $8$. We may assume that $n\equiv 4\ (\text{mod}\ 8)$ because we can replace $\Gamma$ by $\Gamma\bot\Gamma$ or $\Gamma\bot\Gamma\bot\Gamma\bot\Gamma$. Since $\Gamma_C$ is even and unimodular, by Proposition 3 we have
\begin{equation*}
\vartheta_{\Gamma_C}\left(-\frac{1}{z} \right)=\frac{1}{\text{det}(\Gamma_C)} \left(\frac{z}{i}\right)^{\frac{n}{2}} \vartheta_{\Gamma_C^*}(z)=(-1)^{\frac{n}{4}}z^{\frac{n}{2}}\vartheta_{\Gamma_C}(z)=-z^{\frac{n}{2}}\vartheta_{\Gamma_C}(z).
\end{equation*}
Note that $\vartheta_{\Gamma_C}$ is invariant under $T$ based on $\Gamma_C$ is even, i.e.
\begin{equation*}
\vartheta_{\Gamma_C}(z+1)=\sum\limits_{x\in\Gamma_C} e^{\pi i (z+1) x^2}=\sum\limits_{x\in\Gamma_C} e^{\pi i z x^2}=\vartheta_{\Gamma_C}(z),  \tag{2.5}
\end{equation*}
then
\begin{equation*}
\vartheta_{\Gamma_C}((TS)z)=\vartheta_{\Gamma_C}(Sz)=\vartheta_{\Gamma_C}\left(-\frac{1}{z}\right)=-z^{\frac{n}{2}}\vartheta_{\Gamma_C}(z).
\end{equation*}
It follows that
\begin{equation*}
\vartheta_{\Gamma_C}((TS)^3 z)=-((TS)^2 z)^{\frac{n}{2}}\vartheta_{\Gamma_C}((TS)^2 z)=-\left(\frac{1}{1-z}\right)^{\frac{n}{2}} [-((TS)z)^{\frac{n}{2}}] \vartheta_{\Gamma_C}((TS)z)
\end{equation*}
\begin{equation*}
=\left(\frac{1}{1-z}\right)^{\frac{n}{2}} \left(-\frac{1}{z}+1\right)^{\frac{n}{2}}(-z^{\frac{n}{2}})\vartheta_{\Gamma_C}(z)=-\vartheta_{\Gamma_C}(z).\quad\
\end{equation*}
However, this is a contradiction since $(TS)^3=1$. So $n\equiv 0\ (\text{mod}\ 8)$.

To prove $\vartheta_{\Gamma_C}$ is a modular form of weight $\frac{n}{2}$, we already know $\vartheta_{\Gamma_C}$ is holomorphic on $\mathbb{H}\cup \{i\infty\}$, it only needs to show that $\vartheta_{\Gamma_C}(Sz)=z^{\frac{n}{2}} \vartheta_{\Gamma_C}(z)$ and $\vartheta_{\Gamma_C}(Tz)=\vartheta_{\Gamma_C}(z)$ since the modular group $G$ is generated by $S$ and $T$. By Proposition 3, it's easy to verify that
\begin{equation*}
\vartheta_{\Gamma_C}(Sz)=\vartheta_{\Gamma_C}\left(-\frac{1}{z}\right)=\frac{1}{\text{det}(\Gamma_C)}\left(\frac{z}{i}\right)^{\frac{n}{2}}\vartheta_{\Gamma_C^*}(z)=z^{\frac{n}{2}} \vartheta_{\Gamma_C}(z),
\end{equation*}
and we have shown $\vartheta_{\Gamma_C}(Tz)=\vartheta_{\Gamma_C}(z)$ in (2.5). Therefore, $\vartheta_{\Gamma_C}$ is a modular form of weight $\frac{n}{2}$. The proof of Proposition 2 is complete.\\
\hspace*{17.7cm} $\Box$

\section{The Complete Weight Enumerators MacWilliams Identity}

The complete weight enumerators of codes were first proposed by MacWilliams \cite{32} and have been of fundamental importance to theories and practices since they give both the weight enumerators and the frequency of each symbol appearing in each codeword. After that, many researchers extended this work. For example, Wan \cite{44} proved the complete weight enumerators MacWilliams identity for linear codes over Galois ring based on the Fourier transform and Poisson summation formula. In this section, we will provide a proof from the theory of theta function for the complete weight enumerators MacWilliams identity of the codes over $\mathbb{Z}_k$, and obtain the identity for the symmetrized weight enumerators directly.

Let's consider the following $k$ functions.

\textbf{Definition 7} Assume that $\Gamma=\sqrt{k}\mathbb{Z}$ and $t=e^{k\pi i z}$, $z$ is in the upper half plane in $\mathbb{C}$. We define the function
\begin{equation*}
A_0(z)=\sum\limits_{x\in \mathbb{Z}} t^{x^2}=\sum\limits_{x\in \Gamma} t^{\frac{1}{k} x^2}=\sum\limits_{x\in k\mathbb{Z}} t^{\frac{1}{k^2} x^2},
\end{equation*}
which is the theta function of the lattice $\Gamma$. For $1\leqslant j\leqslant k-1$, we define another $k-1$ functions as
\begin{equation*}
A_j(z)=\sum\limits_{x\in k\mathbb{Z}+j} t^{\frac{1}{k^2} x^2}.
\end{equation*}
Note that
\begin{equation*}
\sum\limits_{j=0}^{k-1} A_j(z)=\sum\limits_{x\in \mathbb{Z}} t^{\frac{1}{k^2}x^2}=\sum\limits_{x\in \Gamma^*} t^{\frac{1}{k}x^2},
\end{equation*}
which is the theta function of the dual lattice of $\Gamma$. The following Lemma 1 gives the detailed relationships of these $k$ functions $A_0(z),A_1(z),\cdots,A_{k-1}(z)$.

\textbf{Lemma 1} For any $0\leqslant j\leqslant k-1$,
\begin{equation*}
A_j\left(-\frac{1}{z}\right)=\frac{1}{\sqrt{k}}\left(\frac{z}{i}\right)^{\frac{1}{2}} \sum\limits_{m=0}^{k-1} e^{\frac{2\pi jm}{k}i} A_m(z).
\end{equation*}

\textbf{Proof:} Taking $y=\frac{x-j}{\sqrt{k}}$ in the following equality, we have
\begin{equation*}
A_j(z)=\sum\limits_{x\in k\mathbb{Z}+j} t^{\frac{1}{k^2} x^2}=\sum\limits_{y=\frac{x-j}{\sqrt{k}}\in \Gamma} t^{\frac{1}{k^2} (\sqrt{k}y+j)^2}
\end{equation*}
\begin{equation*}
\quad\ \ =\sum\limits_{y\in \Gamma} t^{\frac{1}{k} (y+\frac{j}{\sqrt{k}})^2}=\sum\limits_{y\in \Gamma} e^{\pi iz(y+\frac{j}{\sqrt{k}})^2}.
\end{equation*}
Therefore,
\begin{equation*}
A_j\left(-\frac{1}{z}\right)=\sum\limits_{y\in \Gamma} e^{-\frac{\pi i}{z}(y+\frac{j}{\sqrt{k}})^2}.
\end{equation*}
Let $f(y)=e^{-\frac{\pi i}{z}(y+\frac{j}{\sqrt{k}})^2}$ be the function of $y$, we calculate the Fourier transform of $f(y)$ and get
\begin{equation*}
\widehat{f}(x)=\int_{\mathbb{R}} f(y) e^{-2\pi i xy}\mathrm{d}y=\left(\frac{z}{i}\right)^{\frac{1}{2}} e^{\frac{2\pi j x}{\sqrt{k}}i} e^{\pi izx^2}.
\end{equation*}
Note that $\Gamma^*=\frac{1}{\sqrt{k}}\mathbb{Z}$, based on the Poisson summation formula,
\begin{equation*}
A_j\left(-\frac{1}{z}\right)=\sum\limits_{y\in \Gamma} e^{-\frac{\pi i}{z}(y+\frac{j}{\sqrt{k}})^2}=\frac{1}{\text{det}(\Gamma)} \left(\frac{z}{i}\right)^{\frac{1}{2}} \sum\limits_{x\in \Gamma^*} e^{\frac{2\pi j x}{\sqrt{k}}i} e^{\pi izx^2}
\end{equation*}
\begin{equation*}
=\frac{1}{\sqrt{k}} \left(\frac{z}{i}\right)^{\frac{1}{2}} \sum\limits_{x\in \mathbb{Z}} e^{\frac{2\pi j x}{k}i} e^{\frac{\pi izx^2}{k}}.\qquad\qquad\quad
\end{equation*}
Since $\mathbb{Z}=\mathop{\bigcup}\limits_{m=0}^{k-1} \{k\mathbb{Z}+m\}$, we have
\begin{equation*}
A_j\left(-\frac{1}{z}\right)=\frac{1}{\sqrt{k}} \left(\frac{z}{i}\right)^{\frac{1}{2}} \sum\limits_{x\in \mathbb{Z}} e^{\frac{2\pi j x}{k}i} e^{\frac{\pi izx^2}{k}}
\end{equation*}
\begin{equation*}
\qquad\qquad\qquad\qquad=\frac{1}{\sqrt{k}} \left(\frac{z}{i}\right)^{\frac{1}{2}} \sum\limits_{m=0}^{k-1} \sum\limits_{x\in k\mathbb{Z}+m} e^{\frac{2\pi j x}{k}i} t^{\frac{1}{k^2}x^2}
\end{equation*}
\begin{equation*}
\qquad\qquad\quad\ \ =\frac{1}{\sqrt{k}} \left(\frac{z}{i}\right)^{\frac{1}{2}} \sum\limits_{m=0}^{k-1} e^{\frac{2\pi jm}{k}i} A_m(z).
\end{equation*}
We complete the proof of Lemma 1.\\
\hspace*{17.7cm} $\Box$

Let $C\subset \mathbb{Z}_k^n$ be a $k$-ary linear code, $\Gamma_C=\frac{1}{\sqrt{k}}\rho^{-1}(C)$ be the associated lattice of $C$. For any $c=(c_1,c_2,\cdots,c_n)\in C$, we denote by $w(c)$ the Hamming weight of $c$, that is,
\begin{equation*}
w(c)=\#\{i\ |\ c_i\neq 0,\ 1\leqslant i\leqslant n\},
\end{equation*}
which is the number of nonzero character in the codeword $c$. For $0\leqslant j\leqslant k-1$, we define $w_j(c)$ as the weight at $j$ of $c$, i.e.
\begin{equation*}
w_j(c)=\#\{i\ |\ c_i=j,\ 1\leqslant i\leqslant n\}.
\end{equation*}
The complete weights of the codeword $c$ are composed of $w_0(c),w_1(c),\cdots,w_{k-1}(c)$, and the complete weight enumerator of $C$ is the polynomial defined as
\begin{equation*}
W_C(X_0,X_1,\cdots,X_{k-1})=\sum\limits_{c\in C} X_0^{w_0(c)} X_1^{w_1(c)}\cdots X_{k-1}^{w_{k-1}(c)}.  \tag{3.1}
\end{equation*}
The degree of the complete weight enumerator is $n$ since $\sum\limits_{j=0}^{k-1} w_j(c)=n$ for all $c\in C$.

On the other hand, we define the symmetrized weight enumerator as
\begin{equation*}
\left\{\begin{array}{ll} S_C(X_0,X_1,\cdots,X_{\frac{k-1}{2}})=\sum\limits_{c\in C} X_0^{w_0(c)} X_1^{w_1(c)+w_{k-1}(c)}\cdots X_{\frac{k-1}{2}}^{w_{\frac{k-1}{2}}(c)+w_{\frac{k+1}{2}}(c)}, & \text{if}\ k\ \text{is odd}, \\ S_C(X_0,X_1,\cdots,X_{\frac{k}{2}})=\sum\limits_{c\in C} X_0^{w_0(c)} X_1^{w_1(c)+w_{k-1}(c)}\cdots X_{\frac{k-2}{2}}^{w_{\frac{k-2}{2}}(c)+w_{\frac{k+2}{2}}(c)} X_{\frac{k}{2}}^{w_{\frac{k}{2}}(c)}, & \text{if}\ k\ \text{is even}. \end{array} \right.
\end{equation*}
Similarly, the degree of the symmetrized weight enumerator is also $n$.

In this section, the main work is to prove Theorem 2 showed in Section I based on the theory of theta function. In order to prove Theorem 2, we first give a few auxiliary lemmas.

\textbf{Lemma 2} Let $C\subset \mathbb{Z}_k^n$ be a $k$-ary code. Then
\begin{equation*}
\text{det}(\rho^{-1}(C))\cdot |C|=k^n.
\end{equation*}

\textbf{Proof:} Since $\rho$ is the natural homomorphism, we have the isomorphism of quotient group $\mathbb{Z}^n/\rho^{-1}(C)\cong \mathbb{Z}_k^n/C$. Hence,
\begin{equation*}
|\mathbb{Z}^n/\rho^{-1}(C)|=|\mathbb{Z}_k^n/C|,
\end{equation*}
this implies that
\begin{equation*}
\text{det}(\rho^{-1}(C))=\text{vol}(\mathbb{R}^n/\rho^{-1}(C))=|\mathbb{Z}^n/\rho^{-1}(C)|\text{vol}(\mathbb{R}^n/\mathbb{Z}^n)=|\mathbb{Z}_k^n/C|=\frac{k^n}{|C|}.
\end{equation*}
Therefore,
\begin{equation*}
\text{det}(\rho^{-1}(C))\cdot |C|=k^n.
\end{equation*}
We finish the proof of Lemma 2.\\
\hspace*{17.7cm} $\Box$

\textbf{Lemma 3} Let $C\subset \mathbb{Z}_k^n$ be a $k$-ary code, $\Gamma_C=\frac{1}{\sqrt{k}}\rho^{-1}(C)$ be the associated lattice of $C$. We have
\begin{equation*}
\Gamma_C^*=\Gamma_{C^{\bot}}.
\end{equation*}

\textbf{Proof:} Note that $\Gamma_C^*=\Gamma_{C^{\bot}}\Leftrightarrow \sqrt{k}\rho^{-1}(C)^*=\frac{1}{\sqrt{k}} \rho^{-1}(C^{\bot})\Leftrightarrow \rho^{-1}(C)^*=\frac{1}{k}\rho^{-1}(C^{\bot})$.

We first prove that $\rho^{-1}(C)^*\subset \frac{1}{k}\rho^{-1}(C^{\bot})$. For any $\alpha\in \rho^{-1}(C)^*$, we note that $c\in \rho^{-1}(C)$ if $c\in C$, it follows that $\alpha\cdot c\in \mathbb{Z}$ for all $c\in C$, then
\begin{equation*}
k\alpha\cdot c\equiv 0\ (\text{mod}\ k),\quad \forall c\in C.
\end{equation*}
This means that $k\alpha\ \text{mod}\ k\in C^{\bot}$, and $k\alpha\in \rho^{-1}(C^{\bot})$, which implies that $\alpha \in \frac{1}{k}\rho^{-1}(C^{\bot})$.

To show that $\frac{1}{k}\rho^{-1}(C^{\bot}) \subset \rho^{-1}(C)^*$, for any $\beta\in \frac{1}{k}\rho^{-1}(C^{\bot})$, or $k\beta\in \rho^{-1}(C^{\bot})$, we prove that $\beta\in \rho^{-1}(C)^*$. For any $c\in C$, we have $(k\beta\ \text{mod}\ k)\cdot c=0$, which implies that $k\beta\cdot c\equiv 0\ (\text{mod}\ k)$. Thus we have $\beta\cdot c\in \mathbb{Z}$ for all $c\in C$. Let $x\in \rho^{-1}(C)$, denote $x\ \text{mod}\ k=c_0\in C$, then we have $\beta\cdot x\in \mathbb{Z}$ since $\beta\cdot c_0\in \mathbb{Z}$, this leads to $\beta\in \rho^{-1}(C)^*$, and $\frac{1}{k}\rho^{-1}(C^{\bot}) \subset \rho^{-1}(C)^*$. We have Lemma 3.\\
\hspace*{17.7cm} $\Box$

Now we can give the proofs of Theorem 1 and Theorem 2.

\textbf{Proof of Theorem 1:} For any $c=(c_1,c_2,\cdots,c_n)\in C$, it follows that
\begin{equation*}
\rho^{-1}(c)=(c_1+k\mathbb{Z})\times (c_2+k\mathbb{Z})\times \cdots \times (c_n+k\mathbb{Z}).
\end{equation*}
It's not difficult to see
\begin{equation*}
\sum\limits_{x\in \frac{1}{\sqrt{k}}\rho^{-1}(c)} t^{\frac{1}{k}x^2}=\sum\limits_{x\in \rho^{-1}(c)} t^{\frac{1}{k^2}x^2}\qquad\qquad\qquad\qquad\qquad\qquad\quad\ \
\end{equation*}
\begin{equation*}
\qquad\qquad\qquad\quad\ \ =\sum\limits_{x_1\in c_1+k\mathbb{Z}} t^{\frac{1}{k^2}x_1^2} \sum\limits_{x_2\in c_2+k\mathbb{Z}} t^{\frac{1}{k^2}x_2^2}\cdots \sum\limits_{x_n\in c_n+k\mathbb{Z}} t^{\frac{1}{k^2}x_n^2}
\end{equation*}
\begin{equation*}
=A_{c_1}(z)A_{c_2}(z)\cdots A_{c_n}(z)\qquad\
\end{equation*}
\begin{equation*}
\qquad\qquad\ \ =A_0(z)^{w_0(c)} A_1(z)^{w_1(c)}\cdots A_{k-1}(z)^{w_{k-1}(c)}.
\end{equation*}
Therefore,
\begin{equation*}
\vartheta_{\Gamma_C}(z)=\sum\limits_{x\in \Gamma_C} t^{\frac{1}{k}x^2}=\sum\limits_{c\in C} \sum\limits_{x\in \frac{1}{\sqrt{k}}\rho^{-1}(c)} t^{\frac{1}{k}x^2}\qquad
\end{equation*}
\begin{equation*}
\qquad\qquad\qquad\ =\sum\limits_{c\in C} A_0(z)^{w_0(c)} A_1(z)^{w_1(c)}\cdots A_{k-1}(z)^{w_{k-1}(c)}
\end{equation*}
\begin{equation*}
\qquad=W_C(A_0(z),A_1(z),\cdots,A_{k-1}(z)).
\end{equation*}
This is the proof of Theorem 1.\\
\hspace*{17.7cm} $\Box$

\textbf{Proof of Theorem 2:} From Theorem 1, Lemma 2, Lemma 3 and Proposition 2, one can get
\begin{equation*}
W_C\left(A_0\left(-\frac{1}{z}\right),A_1\left(-\frac{1}{z}\right),\cdots,A_{k-1}\left(-\frac{1}{z}\right)\right)=\vartheta_{\Gamma_C}\left(-\frac{1}{z}\right)
\end{equation*}
\begin{equation*}
=\frac{1}{\text{det}(\Gamma_C)} \left(\frac{z}{i}\right)^{\frac{n}{2}} \vartheta_{\Gamma_C^*}(z)=\frac{1}{\text{det}(\rho^{-1}(C))/k^{\frac{n}{2}}} \left(\frac{z}{i}\right)^{\frac{n}{2}} \vartheta_{\Gamma_{C^{\bot}}}(z)\qquad\ \
\end{equation*}
\begin{equation*}
=\frac{|C|}{k^{\frac{n}{2}}} \left(\frac{z}{i}\right)^{\frac{n}{2}} W_{C^{\bot}}(A_0(z),A_1(z),\cdots,A_{k-1}(z)).\qquad\qquad\qquad\qquad  \tag{3.2}
\end{equation*}
On the other hand, since the complete weight enumerator $W_C$ is a homogeneous polynomial of degree $n$, it implies that
\begin{equation*}
W_C\left(A_0\left(-\frac{1}{z}\right),A_1\left(-\frac{1}{z}\right),\cdots,A_{k-1}\left(-\frac{1}{z}\right)\right)
\end{equation*}
\begin{equation*}
=\frac{1}{k^{\frac{n}{2}}} \left(\frac{z}{i}\right)^{\frac{n}{2}} W_C(\sum\limits_{j=0}^{k-1} A_j(z), \sum\limits_{j=0}^{k-1} e^{\frac{2\pi j}{k}i}A_j(z), \cdots, \sum\limits_{j=0}^{k-1} e^{\frac{2\pi(k-1) j}{k}i}A_j(z)).  \tag{3.3}
\end{equation*}
Comparing with (3.2) and (3.3), we have
\begin{equation*}
W_{C^{\bot}}(A_0(z),A_1(z),\cdots,A_{k-1}(z))=\frac{1}{|C|} W_C(\sum\limits_{j=0}^{k-1} A_j(z), \sum\limits_{j=0}^{k-1} e^{\frac{2\pi j}{k}i}A_j(z), \cdots, \sum\limits_{j=0}^{k-1} e^{\frac{2\pi(k-1) j}{k}i}A_j(z)).
\end{equation*}
We finish the proof of Theorem 2.\\
\hspace*{17.7cm} $\Box$

\textbf{Corollary 1} Let $C\subset \mathbb{Z}_k^n$ be a $k$-ary self-dual code, then
\begin{equation*}
W_C(A_0(z),A_1(z),\cdots,A_{k-1}(z))=W_C\left(\frac{1}{\sqrt{k}} \sum\limits_{j=0}^{k-1} A_j(z), \frac{1}{\sqrt{k}} \sum\limits_{j=0}^{k-1} e^{\frac{2\pi j}{k}i}A_j(z), \cdots, \frac{1}{\sqrt{k}} \sum\limits_{j=0}^{k-1} e^{\frac{2\pi(k-1) j}{k}i}A_j(z)\right).
\end{equation*}

\textbf{Proof:} From $C$ is a self-dual code we can get $|C|=k^{\frac{n}{2}}=\sqrt{k}^n$. Note that $W_C$ is a homogeneous polynomial of degree $n$, by Theorem 2 we have
\begin{equation*}
W_C(A_0(z),A_1(z),\cdots,A_{k-1}(z))=W_{C^{\bot}} (A_0(z),A_1(z),\cdots,A_{k-1}(z))
\end{equation*}
\begin{equation*}
=\frac{1}{\sqrt{k}^n} W_C(\sum\limits_{j=0}^{k-1} A_j(z), \sum\limits_{j=0}^{k-1} e^{\frac{2\pi j}{k}i}A_j(z), \cdots, \sum\limits_{j=0}^{k-1} e^{\frac{2\pi(k-1) j}{k}i}A_j(z))\quad
\end{equation*}
\begin{equation*}
\qquad\quad\ \ =W_C\left(\frac{1}{\sqrt{k}} \sum\limits_{j=0}^{k-1} A_j(z), \frac{1}{\sqrt{k}} \sum\limits_{j=0}^{k-1} e^{\frac{2\pi j}{k}i}A_j(z), \cdots, \frac{1}{\sqrt{k}} \sum\limits_{j=0}^{k-1} e^{\frac{2\pi(k-1) j}{k}i}A_j(z)\right).
\end{equation*}
\hspace*{17.7cm} $\Box$

Corollary 1 shows that the complete weight enumerator of a self-dual code in $\mathbb{Z}_k^n$ is invariant under a rotation in $\mathbb{R}^n$. The following Corollary 2 provides the symmetrized weight enumerators MacWilliams identity for codes over $\mathbb{Z}_k$.

\textbf{Corollary 2} Let $C\subset \mathbb{Z}_k^n$ be a $k$-ary code, then we have
\begin{equation*}
S_{C^{\bot}} (A_0,A_1,\cdots,A_{\frac{k-1}{2}})=\frac{1}{|C|} S_C(A_0+\sum\limits_{j=1}^{\frac{k-1}{2}} 2A_j,A_0+\sum\limits_{j=1}^{\frac{k-1}{2}} 2\cos\frac{2j\pi}{k} A_j,\cdots,A_0+\sum\limits_{j=1}^{\frac{k-1}{2}} 2\cos\frac{(k-1)j\pi}{k} A_j)
\end{equation*}
if $k$ is odd, and
\begin{equation*}
S_{C^{\bot}} (A_0,A_1,\cdots,A_{\frac{k}{2}})=\frac{1}{|C|} S_C(A_0+\sum\limits_{j=1}^{\frac{k-2}{2}} 2A_j+A_{\frac{k}{2}},A_0+\sum\limits_{j=1}^{\frac{k-2}{2}} 2\cos\frac{2j\pi}{k} A_j-A_{\frac{k}{2}},\cdots,
\end{equation*}
\begin{equation*}
A_0+\sum\limits_{j=1}^{\frac{k-2}{2}} 2\cos\frac{(k-2)j\pi}{k} A_j+(-1)^{\frac{k-2}{2}} A_{\frac{k}{2}},A_0+\sum\limits_{j=1}^{\frac{k-2}{2}}2(-1)^{j}A_j+(-1)^{\frac{k}{2}}A_{\frac{k}{2}})
\end{equation*}
if $k$ is even.

\textbf{Proof:} If $k$ is odd, based on the definition of the symmetrized weight enumerator, we have
\begin{equation*}
S_{C^{\bot}} (A_0,A_1,\cdots,A_{\frac{k-1}{2}})=\sum\limits_{c\in C^{\bot}} A_0^{w_0(c)} A_1^{w_1(c)+w_{k-1}(c)}\cdots A_{\frac{k-1}{2}}^{w_{\frac{k-1}{2}}(c)+w_{\frac{k+1}{2}}(c)}
\end{equation*}
\begin{equation*}
\qquad\qquad\qquad\qquad\qquad\ =W_{C^{\bot}}(A_0,A_1,A_2,\cdots,A_{\frac{k-1}{2}},A_{\frac{k-1}{2}},\cdots,A_2,A_1)
\end{equation*}
\begin{equation*}
\qquad\qquad\quad=\frac{1}{|C|}W_C(B_0,B_1,B_2,\cdots,B_{k-2},B_{k-1}),  \tag{3.4}
\end{equation*}
where $B_0,B_1,B_2,\cdots,B_{k-2},B_{k-1}$ satisfy that
\begin{equation*}
B_0=A_0+\sum\limits_{j=1}^{\frac{k-1}{2}} 2A_j,
\end{equation*}
\begin{equation*}
B_1=B_{k-1}=A_0+\sum\limits_{j=1}^{\frac{k-1}{2}} 2\cos\frac{2j\pi}{k} A_j,\cdots
\end{equation*}
\begin{equation*}
B_{\frac{k-1}{2}}=B_{\frac{k+1}{2}}=A_0+\sum\limits_{j=1}^{\frac{k-1}{2}} 2\cos\frac{(k-1)j\pi}{k} A_j.
\end{equation*}
Therefore, from (3.4) we get
\begin{equation*}
S_{C^{\bot}} (A_0,A_1,\cdots,A_{\frac{k-1}{2}})=\frac{1}{|C|}W_C(B_0,B_1,B_2,\cdots,B_{k-2},B_{k-1})=\frac{1}{|C|}S_C(B_0,B_1,\cdots,B_{\frac{k-1}{2}})
\end{equation*}
\begin{equation*}
=\frac{1}{|C|} S_C(A_0+\sum\limits_{j=1}^{\frac{k-1}{2}} 2A_j,A_0+\sum\limits_{j=1}^{\frac{k-1}{2}} 2\cos\frac{2j\pi}{k} A_j,\cdots,A_0+\sum\limits_{j=1}^{\frac{k-1}{2}} 2\cos\frac{(k-1)j\pi}{k} A_j).
\end{equation*}
If $k$ is even, we can get the identity in the same way.\\
\hspace*{17.7cm} $\Box$

\section{The Complete Weight Enumerators MacWilliams Identity in Genus $g$}

In this section, we present the complete weight enumerators MacWilliams identity in genus $g$, which is a generalization of the work by Bannai et al. \cite{02} to the finite ring $\mathbb{Z}_k$. First let's introduce some definitions and notations.

\textbf{Definition 8} Let $C\subset \mathbb{Z}_k^n$ be a $k$-ary code. For a positive integer $g$, we define the complete weight enumerators in genus $g$ of the code $C$ is
\begin{equation*}
\mathfrak{C}_{C,g}(z_a\ \text{with}\ a\in \mathbb{Z}_k^g)=\sum\limits_{c_1,\cdots,c_g\in C} \mathop{\prod}\limits_{a\in \mathbb{Z}_k^g} z_a^{w_a(c_1,\cdots,c_g)},  \tag{4.1}
\end{equation*}
if we write $c_1=(c_{11},c_{12},\cdots,c_{1n}),\cdots,c_g=(c_{g1},c_{g2},\cdots,c_{gn})$, here $w_a(c_1,\cdots,c_g)$ denotes the number of $i$ satisfying $(c_{1i},c_{2i},\cdots,c_{gi})=a$, i.e.
\begin{equation*}
w_a(c_1,\cdots,c_g)=\#\{i\ |\ (c_{1i},c_{2i},\cdots,c_{gi})=a,\ 1\leqslant i\leqslant n\}.
\end{equation*}
It's easy to see that if $g=1$, then the above complete weight enumerators in genus $1$ is the same as the complete weight enumerators (3.1) defined in the previous section.

Assume $f\in \mathbb{C}(x_1,x_2,\cdots,x_n)$ is a complex polynomial of $x_1,x_2,\cdots,x_n$, and $M$ is a matrix $(a_{ij})_{n\times n}$ of order $n$. We denote $Mf(x_1,x_2,\cdots,x_n)$ by
\begin{equation*}
Mf(x_1,x_2,\cdots,x_n)=f\left(\sum\limits_{j=1}^n a_{1j}x_j,\sum\limits_{j=1}^n a_{2j}x_j,\cdots,\sum\limits_{j=1}^n a_{nj}x_j\right).   \tag{4.2}
\end{equation*}

In \cite{02}, Bannai et al. give the following MacWilliams identity of the complete weight enumerators in Genus $g$ for an even number $k$:
\begin{equation*}
\mathfrak{C}_{C^{\bot},g}(z_a)=\frac{1}{|C|^g} T\mathfrak{C}_{C,g}(z_a),
\end{equation*}
here $T=(\eta^{a\cdot b})_{a,b\in \mathbb{Z}_k^g}$, and $\eta=e^{\frac{2\pi i}{k}}$ is the primitive root of unit. We set an additive characteristic $\psi$ for $\mathbb{Z}_k$ by $\psi(x)=e^{2\pi ix/k}$, $x\in \mathbb{Z}_k$. Now we give a proof of Theorem 3 in Section I and show that it also holds for codes $C\subset \mathbb{Z}_k^n$ for any positive integer $k$.

\textbf{Proof of Theorem 3:} For any $c_1,c_2,\cdots,c_g\in C$, we write $c_1=(c_{11},c_{12},\cdots,c_{1n}),\cdots,c_g=(c_{g1},c_{g2},\cdots,c_{gn})$. Let the function $f$ be
\begin{equation*}
f(c_1,c_2,\cdots,c_g)=\mathop{\prod}\limits_{a\in \mathbb{Z}_k^g} z_a^{w_a(c_1,\cdots,c_g)},
\end{equation*}
here $w_a(c_1,\cdots,c_g)$ is defined in (4.2). To prove Theorem 3, we first show the Fourier transform of $f(c_1,c_2,\cdots,c_g)$ is
\begin{equation*}
\widehat{f}(c_1,c_2,\cdots,c_g)=Tf(c_1,c_2,\cdots,c_g).
\end{equation*}
We denote $c$ by the $g\times n$ matrix $(c_{ij})_{g\times n}$ composed of $c_1,c_2,\cdots,c_g$ as the $g$ rows, and denote $x_1,x_2,\cdots,x_n$ by the $n$ columns of the matrix $c$. For any $\xi_1,\xi_2,\cdots,\xi_g\in \mathbb{Z}_k^n$, let $\xi$ be the matrix composed of $\xi_1,\xi_2,\cdots,\xi_g$ as the rows, and $y_1,y_2,\cdots,y_n$ be the columns of $\xi$. It follows that the Fourier transform of $f(c_1,c_2,\cdots,c_g)$ is given by
\begin{equation*}
\widehat{f}(c_1,c_2,\cdots,c_g)=\sum\limits_{\xi_1,\cdots,\xi_g\in \mathbb{Z}_k^n} (\mathop{\prod}\limits_{a\in \mathbb{Z}_k^g} z_a^{w_a(\xi_1,\cdots,\xi_g)}) \psi(<c,\xi>),
\end{equation*}
here $<c,\xi>$ is the trace of $c^T \xi$, which is equal to $\sum\limits_{j=1}^n x_i\cdot y_i$. Note that $w_a(\xi_1,\xi_2,\cdots,\xi_g)=\sum\limits_{j=1}^n w_a(y_j)$, and
\begin{equation*}
\psi(<c,\xi>)=\psi(\sum\limits_{j=1}^n x_j\cdot y_j)=\mathop{\prod}\limits_{j=1}^n \psi(x_j\cdot y_j),
\end{equation*}
therefore, we have
\begin{equation*}
\widehat{f}(c_1,c_2,\cdots,c_g)=\sum\limits_{\xi_1,\cdots,\xi_g\in \mathbb{Z}_k^n} (\mathop{\prod}\limits_{a\in \mathbb{Z}_k^g} z_a^{w_a(\xi_1,\cdots,\xi_g)}) \psi(<c,\xi>)
\end{equation*}
\begin{equation*}
\qquad\qquad\qquad\qquad\qquad=\sum\limits_{\xi=(y_1,\cdots,y_n)\in \mathbb{Z}_k^{g\times n}}  \mathop{\prod}\limits_{j=1}^n \left( \psi(x_j\cdot y_j) \mathop{\prod}\limits_{a\in \mathbb{Z}_k^g} z_a^{w_a(y_j)} \right)
\end{equation*}
\begin{equation*}
\qquad\qquad\quad\ =\mathop{\prod}\limits_{j=1}^n \sum\limits_{y_j\in \mathbb{Z}_k^g} \left( \psi(x_j\cdot y_j) \mathop{\prod}\limits_{a\in \mathbb{Z}_k^g}  z_a^{w_a(y_j)}\right)
\end{equation*}
\begin{equation*}
=\mathop{\prod}\limits_{j=1}^n \sum\limits_{a\in \mathbb{Z}_k^g} \psi(x_j\cdot a)z_a\quad\
\end{equation*}
\begin{equation*}
\qquad\qquad=\mathop{\prod}\limits_{b\in \mathbb{Z}_k^g} (\sum\limits_{a\in \mathbb{Z}_k^g}\psi(b\cdot a)z_a )^{w_b(c_1,\cdots,c_g)}.
\end{equation*}
Based on the Poisson summation formula,
\begin{equation*}
\sum\limits_{c_1,\cdots,c_g\in C^{\bot}}f(c_1,c_2,\cdots,c_g)=\frac{1}{|C|^g}\sum\limits_{c_1,\cdots,c_g\in C} \widehat{f}(c_1,c_2,\cdots,c_g).
\end{equation*}
Combine with (4.1) and (4.2), one can get
\begin{equation*}
\mathfrak{C}_{C^{\bot},g}(z_a)=\sum\limits_{c_1,\cdots,c_g\in C^{\bot}}f(c_1,c_2,\cdots,c_g)\qquad\qquad\qquad\qquad\qquad\qquad
\end{equation*}
\begin{equation*}
\qquad\qquad\qquad\qquad\ \ =\frac{1}{|C|^g} \sum\limits_{c_1,\cdots,c_g\in C} \mathop{\prod}\limits_{b\in \mathbb{Z}_k^g} (\sum\limits_{a\in \mathbb{Z}_k^g}\psi(b\cdot a)z_a )^{w_b(c_1,\cdots,c_g)}=\frac{1}{|C|^g} T\mathfrak{C}_{C,g}(z_a).
\end{equation*}
This is the proof of Theorem 3.\\
\hspace*{17.7cm} $\Box$.

\section{The Complete Weight Enumerators MacWilliams Identity in Cyclotomic Fields}

In this section, we assume that $k=p$ is an odd prime number and $C\subset F_p^n$ is a linear code over $F_p$ satisfying $C\subset C^{\bot}$. Let $\xi=e^{\frac{2\pi i}{p}}$ and $K=\mathbb{Q}(\xi)$ be the cyclotomic field obtained by adjoining $\xi$ to $\mathbb{Q}$. Since $p-1$ is the degree of the minimal polynomial of $\xi$ over $\mathbb{Q}$, then $K$ is a vector space over $\mathbb{Q}$ of dimension $p-1$. Assume that $K^+=\mathbb{Q}(\xi+\xi^{-1})$ is the maximal real subfield of $K$. We denote $\text{Tr}_{K/\mathbb{Q}}$ by the trace function of elements in $K$. Let $\mathfrak{D}$ be the ring of integers of $K$, i.e.
\begin{equation*}
\mathfrak{D}=\{\alpha=\sum\limits_{j=0}^{p-2}a_j \xi^j\ |\ a_j\in \mathbb{Z},\ j=0,1,\cdots,p-2\}.
\end{equation*}
Suppose $\mathfrak{B}=\left<1-\xi\right>$ is the principal ideal of $\mathfrak{D}$ generated by the element $1-\xi\in \mathfrak{D}$. We define $\rho: \mathfrak{D}^n\longrightarrow (\mathfrak{D}/\mathfrak{B})^n$ by the mapping of the reduction modulo the principal ideal $\mathfrak{B}$ in each coordinate. For example, if $\alpha=\sum\limits_{j=0}^{p-2}a_j \xi^j\in \mathfrak{D}$, $a_0,a_1,\cdots,a_{p-2}\in \mathbb{Z}$, it's not hard to get
\begin{equation*}
\rho(\alpha)= \sum\limits_{j=0}^{p-2}a_j\ \text{mod}\ p,
\end{equation*}
which indicates that $\mathfrak{D}/\mathfrak{B}\cong F_p$. For any $x=(x_1,x_2,\cdots,x_n)\in \mathfrak{D}$, $y=(y_1,y_2,\cdots,y_n)\in \mathfrak{D}$, we define $x\overline{y}=\sum\limits_{j=1}^n x_j \overline{y_j}$ where $\overline{y}$ is the complex conjugate of $y$. Let $\Gamma_C=\rho^{-1}(C)\subset \mathfrak{D}^n$ be the associated lattice of the code $C$. The theta function of $\Gamma_C$ is defined as
\begin{equation*}
\vartheta_{\Gamma_C}(z)=\sum\limits_{x\in \Gamma_C}e^{2\pi iz \text{Tr}_{K^+/\mathbb{Q}}(\frac{x \overline{x}}{p})},
\end{equation*}
where $z\in \mathbb{H}$ is the upper half plane of complex number. For any $j=0,1,2,\cdots,p-1$, suppose
\begin{equation*}
\vartheta_j(z)=\sum\limits_{x\in \mathfrak{B}+j} e^{2\pi iz \text{Tr}_{K^+/\mathbb{Q}}(\frac{x \overline{x}}{p})}.
\end{equation*}

Now we give the proof of Theorem 4 given in Section I.

\textbf{Proof of Theorem 4:} For any $c=(c_1,c_2,\cdots,c_n)\in C$, it follows that
\begin{equation*}
\rho^{-1}(c)=(c_1+\mathfrak{B})\times (c_2+\mathfrak{B})\times \cdots \times (c_n+\mathfrak{B}).
\end{equation*}
Then we have
\begin{equation*}
\sum\limits_{x\in \rho^{-1}(c)} e^{2\pi iz \text{Tr}_{K^+/\mathbb{Q}}(\frac{x\overline{x}}{p})}=\sum\limits_{x=(x_1,\cdots,x_n)\in \rho^{-1}(c)} e^{2\pi iz \text{Tr}_{K^+/\mathbb{Q}}(\frac{x_1 \overline{x_1}+x_2\overline{x_2}+\cdots+x_n\overline{x_n}}{p})}
\end{equation*}
\begin{equation*}
=\sum\limits_{x_1\in c_1+\mathfrak{B}} e^{2\pi iz \text{Tr}_{K^+/\mathbb{Q}}(\frac{x_1\overline{x_1}}{p})} \sum\limits_{x_2\in c_2+\mathfrak{B}} e^{2\pi iz \text{Tr}_{K^+/\mathbb{Q}}(\frac{x_2\overline{x_2}}{p})} \sum\limits_{x_n\in c_n+\mathfrak{B}} e^{2\pi iz \text{Tr}_{K^+/\mathbb{Q}}(\frac{x_n\overline{x_n}}{p})}
\end{equation*}
\begin{equation*}
=\vartheta_{c_1}(z)\vartheta_{c_2}(z)\cdots \vartheta_{c_n}(z)=\vartheta_0(z)^{w_0(c)} \vartheta_1(z)^{w_1(c)}(z) \vartheta_{p-1}(z)^{w_{p-1}(c)}.\qquad\qquad\qquad\
\end{equation*}
Therefore,
\begin{equation*}
\sum\limits_{c\in C} \sum\limits_{x\in \rho^{-1}(c)} e^{2\pi iz \text{Tr}_{K^+/\mathbb{Q}}(\frac{x\overline{x}}{p})}=\sum\limits_{c\in C} \vartheta_0(z)^{w_0(c)} \vartheta_1(z)^{w_1(c)}(z) \vartheta_{p-1}(z)^{w_{p-1}(c)},
\end{equation*}
which means that
\begin{equation*}
\vartheta_{\Gamma_C}(z)=W_C(\vartheta_0(z),\vartheta_1(z),\cdots,\vartheta_{p-1}(z)).
\end{equation*}
So we finish the proof of Theorem 4.\\
\hspace*{17.7cm} $\Box$.

The similar method may yield a high dimensional result. Let $z=(z_1,z_2,\cdots,z_{p-1})\in \mathbb{H}^{p-1}$. We define the trace and theta function by
\begin{equation*}
\text{Tr}_{K/\mathbb{Q}}(z\frac{x\overline{y}}{p})=\sum\limits_{i=1}^{p-1} z_i \sigma_i(\frac{x \overline{y}}{p}),
\end{equation*}
where $x,y\in \mathfrak{D}$, each $\sigma_i$ is the embedding of $K\longrightarrow \mathbb{C}$,
\begin{equation*}
\theta_j(z)=\sum\limits_{x\in \mathfrak{B}+j} e^{\pi i \text{Tr}_{K/\mathbb{Q}}(z\frac{x\overline{x}}{p})}=\sum\limits_{x\in \mathfrak{B}+j} e^{2\pi i \text{Tr}_{K^+/\mathbb{Q}}(z\frac{x\overline{x}}{p})},\ 0\leqslant j\leqslant p-1,
\end{equation*}
and
\begin{equation*}
\theta_{\Gamma_C}(z)=\sum\limits_{x\in \Gamma_C} e^{\pi i \text{Tr}_{K/\mathbb{Q}}(z\frac{x\overline{x}}{p})}=\sum\limits_{x\in \Gamma_C} e^{2\pi i \text{Tr}_{K^+/\mathbb{Q}}(z\frac{x\overline{x}}{p})},\ z\in \mathbb{H}^{p-1}.
\end{equation*}
We also have
\begin{equation*}
\theta_{\Gamma_C}(z)=W_C(\theta_0(z),\theta_1(z),\cdots,\theta_{p-1}(z)),\ z\in \mathbb{H}^{p-1}.  \tag{5.1}
\end{equation*}

The above result may compare with Theorem 5.1 of \cite{a}, which gave a complete weight enumerator MacWilliams identity with $z\in \mathbb{H}^{\frac{p-1}{2}}$ in the totally real field $\mathbb{Q}(\xi+\xi^{-1})$.

\section{Conclusion}

To show that Theorem 2 is equivalent to the complete weight enumerators MacWilliams identity, it is important to prove the algebraic independence of $A_0,A_1,\cdots,A_{k-1}$. We wish to find a proof based on the algebra of Hilbert modular form. Another topic of this paper is the connection between weight enumerators of codes and theta functions of lattices. The results of van der Geer and Hirzebruch could be considered as a generalization of that on weight enumerators of codes and theta functions of lattices in the binary case. They showed the relationship between the Lee weight enumerators of $p$-ary codes when $p$ is an odd prime number and associated lattices over the ring of algebraic integers on a cyclotomic field. To establish the properties of the complete weight enumerators for codes over $F_p$ with theta functions, we generalize their works. It's also interesting to obtain a generalization for a general positive integer $k$ instead of a prime number $p$. We will discuss this in our future works.

In this paper, we prove the complete and symmetrized weight enumerators MacWilliams identity for codes over $\mathbb{Z}_k$ based on theta functions, and present the complete weight enumerators MacWilliams identity in genus $g$ in general, which is a generalization of the works by Bannai et al. The further questions are to consider the modified theta functions or the nu-function of a lattice associated with a $k$-ary code. Some results could be fould in our previous work \cite{51}. It's interesting to explore whether these functions are a kind of special modular forms, as well as present the MacWilliams identities of these functions based on the theory of theta function and modular form.

\section*{Acknowledgments}

We discuss with professor P. Sol\'{e} about this paper during his visiting to Great Bay University. We would like to thank him for his valuable suggestions. We thank professor Zhedanov for revising the English writing of this paper. This work was supported by Information Security School-Enterprise Joint Laboratory (Dongguan Institute for Advanced Study, Greater Bay Area) and Major Project of Henan Province (No. 225200810036).

\end{document}